\documentclass[twocolumn,showpacs,preprintnumbers,amsmath,amssymb,superscriptaddress]{revtex4}

\usepackage{graphicx}
\usepackage{dcolumn}
\usepackage{bm}

\begin{document}


\title{$\rho^0$ Production and Possible Modification in Au+Au and $p+p$
Collisions at $\sqrt{s_{_{NN}}}$ \!=\! 200 GeV}

\date{\today}

\affiliation{Argonne National Laboratory, Argonne, Illinois 60439}
\affiliation{Brookhaven National Laboratory, Upton, New York
11973} \affiliation{University of Birmingham, Birmingham, United
Kingdom} \affiliation{University of California, Berkeley,
California 94720} \affiliation{University of California, Davis,
California 95616} \affiliation{University of California, Los
Angeles, California 90095} \affiliation{California Institute of
Technology, Pasadena, California 91125} \affiliation{Carnegie
Mellon University, Pittsburgh, Pennsylvania 15213}
\affiliation{Creighton University, Omaha, Nebraska 68178}
\affiliation{Nuclear Physics Institute AS CR, \v{R}e\v{z}/Prague,
Czech Republic} \affiliation{Laboratory for High Energy (JINR),
Dubna, Russia} \affiliation{Particle Physics Laboratory (JINR),
Dubna, Russia} \affiliation{University of Frankfurt, Frankfurt,
Germany} \affiliation{Indian Institute of Technology, Mumbai,
India} \affiliation{Indiana University, Bloomington, Indiana
47408} \affiliation{Institute  of Physics, Bhubaneswar 751005,
India} \affiliation{Institut de Recherches Subatomiques,
Strasbourg, France} \affiliation{University of Jammu, Jammu
180001, India} \affiliation{Kent State University, Kent, Ohio
44242} \affiliation{Lawrence Berkeley National Laboratory,
Berkeley, California 94720}\affiliation{Max-Planck-Institut f\"ur
Physik, Munich, Germany} \affiliation{Michigan State University,
East Lansing, Michigan 48824} \affiliation{Moscow Engineering
Physics Institute, Moscow Russia} \affiliation{City College of New
York, New York City, New York 10031} \affiliation{NIKHEF,
Amsterdam, The Netherlands} \affiliation{Ohio State University,
Columbus, Ohio 43210} \affiliation{Panjab University, Chandigarh
160014, India} \affiliation{Pennsylvania State University,
University Park, Pennsylvania 16802} \affiliation{Institute of
High Energy Physics, Protvino, Russia} \affiliation{Purdue
University, West Lafayette, Indiana 47907} \affiliation{University
of Rajasthan, Jaipur 302004, India} \affiliation{Rice University,
Houston, Texas 77251} \affiliation{Universidade de Sao Paulo, Sao
Paulo, Brazil} \affiliation{University of Science \& Technology of
China, Anhui 230027, China} \affiliation{Shanghai Institute of
Nuclear Research, Shanghai 201800, P.R. China}
\affiliation{SUBATECH, Nantes, France} \affiliation{Texas A\&M
University, College Station, Texas 77843} \affiliation{University
of Texas, Austin, Texas 78712} \affiliation{Valparaiso University,
Valparaiso, Indiana 46383} \affiliation{Variable Energy Cyclotron
Centre, Kolkata 700064, India} \affiliation{Warsaw University of
Technology, Warsaw, Poland} \affiliation{University of Washington,
Seattle, Washington 98195} \affiliation{Wayne State University,
Detroit, Michigan 48201} \affiliation{Institute of Particle
Physics, CCNU (HZNU), Wuhan, 430079 China} \affiliation{Yale
University, New Haven, Connecticut 06520} \affiliation{University
of Zagreb, Zagreb, HR-10002, Croatia}
\author{J.~Adams}\affiliation{University of Birmingham, Birmingham, United Kingdom}
\author{C.~Adler}\affiliation{University of Frankfurt, Frankfurt, Germany}
\author{M.M.~Aggarwal}\affiliation{Panjab University, Chandigarh 160014, India}
\author{Z.~Ahammed}\affiliation{Variable Energy Cyclotron Centre, Kolkata 700064, India}
\author{J.~Amonett}\affiliation{Kent State University, Kent, Ohio 44242}
\author{B.D.~Anderson}\affiliation{Kent State University, Kent, Ohio 44242}
\author{D.~Arkhipkin}\affiliation{Particle Physics Laboratory (JINR), Dubna, Russia}
\author{G.S.~Averichev}\affiliation{Laboratory for High Energy (JINR), Dubna, Russia}
\author{S.K.~Badyal}\affiliation{University of Jammu, Jammu 180001, India}
\author{J.~Balewski}\affiliation{Indiana University, Bloomington, Indiana 47408}
% ...


\begin{abstract}
We report results on $\rho(770)^0 \!\rightarrow\! \pi^+\pi^-$
production at midrapidity in $p+p$ and peripheral Au+Au collisions
at $\sqrt{s_{_{NN}}} \!=\!$ 200 GeV. This is the first direct
measurement of $\rho(770)^0 \!\rightarrow\! \pi^+\pi^-$ in
heavy-ion collisions. The measured $\rho^0$ peak in the invariant
mass distribution is shifted by $\sim$40 MeV/$c^2$ in minimum bias
$p+p$ interactions and $\sim$70 MeV/$c^2$ in peripheral Au+Au
collisions. The $\rho^0$ mass shift is dependent on transverse
momentum and multiplicity. The modification of the $\rho^0$ meson
mass, width, and shape due to phase space and dynamical effects
are discussed.
\end{abstract}


\maketitle

In-medium modification of the $\rho$ meson due to the
effects of increasing temperature and density has been proposed as
a possible signal of a phase transition of nuclear matter to a
deconfined plasma of quarks and gluons, which is expected to be
accompanied by the restoration of chiral symmetry \cite{1}.

The $\rho^0$ meson measured in the dilepton channel probes all
stages of the system formed in relativistic heavy-ion collisions
because the dileptons have negligible final state interactions
with the hadronic environment. Heavy-ion experiments at CERN
indicate an enhanced dilepton production cross section in the
invariant mass range of 200-600 MeV$/c^2$ \cite{2}.  The study of
the dilepton decay channel currently relies on model calculations
based on a so-called ``hadronic cocktail", a superposition of the
expected contributions to the dilepton spectrum \cite{1,2,3}. The
present hadronic decay measurement, $\rho(770)^0 \!\rightarrow\!
\pi^{+}\pi^{-}$, is the first of its kind in heavy-ion collisions
and provides experimental data to help constrain the input to the
hadronic cocktail used for such studies.

Even in the absence of the phase transition, at normal nuclear
density, temperature and density dependent modifications of the
$\rho^0$ meson are expected to be measurable. Effects such as
phase space \cite{4,5,6,7,8,9,10,33,11} and dynamical interactions
with matter \cite{6,8,10} may modify the $\rho^0$ mass, width, and
shape. These modifications of the $\rho^0$ properties take place
close to kinetic freeze-out (vanishing elastic collisions), in a
dilute hadronic gas at late stages of heavy-ion collisions. At
such low matter density, the proposed modifications are expected
to be small, but observable. The effects of phase space due to the
rescattering of pions, $\pi^{+}\pi^{-} \!\rightarrow\! \rho^0
\!\rightarrow\! \pi^{+}\pi^{-}$, and Bose-Einstein correlations
between pions from $\rho^0$ decay and pions in the surrounding
matter are present in $p+p$ \cite{5,6,8,11} and Au+Au
\cite{4,6,7,8,9,10,33} collisions. The interference between
different pion scattering channels can effectively distort the
line shape of resonances \cite{12}. Dynamical effects due to the
$\rho^0$ interacting with the surrounding matter are also expected
to be present in $p+p$ and Au+Au interactions, and have been
evaluated for the latter \cite{6,8,10}.

Since the $\rho^0$ lifetime of $c\tau$ \!=\! 1.3 fm is small with
respect to the lifetime of the system formed in Au+Au collisions,
the $\rho^0$ meson is expected to decay, regenerate, and rescatter
all the way through kinetic freeze-out. In the context of
statistical models, the measured $\rho^0$ yield should reflect
conditions at kinetic freeze-out rather than at chemical
freeze-out (vanishing inelastic collisions) \cite{6,8,9,33,13}. In
$p+p$ collisions, the $\rho^0$ meson is expected to be produced
predominantly by string fragmentation. The measurement of the
$\rho^0$ meson in $p+p$ and Au+Au interactions at the same
nucleon-nucleon c.m. system energy can provide insight for
understanding the dynamics of these systems.

The detector system used for these studies was the Solenoidal
Tracker at RHIC (STAR). The main tracking device within STAR is
the time projection chamber (TPC) \cite{14} located inside a 0.5 T
solenoidal magnetic field. In addition to providing momentum
information, the TPC provides particle identification for charged
particles by measuring their ionization energy loss ($dE/dx$). In
Au+Au collisions, a minimum bias trigger was defined using
coincidences between two zero degree calorimeters that measured
the spectator neutrons. In $p+p$ collisions, the minimum bias
trigger was defined using coincidences between two beam-beam
counters that measured the charged particle multiplicity in
forward pseudorapidities (3.3 $\!<\! |\eta| \!<\!$ 5.0). This
trigger is sensitive to nonsingly diffractive (NSD) events, with
negligible bias on yields \cite{15}. Approximately $11 \times
10^6$ minimum bias $p+p$ events, $1.5 \times 10^6$ high
multiplicity $p+p$ events, and $1.2 \times 10^6$ events in the
peripheral centrality class corresponding to 40-80$\%$ of the
inelastic hadronic Au+Au cross section were used for this
analysis. The beam energy was $\sqrt{s_{_{NN}}}$ \!=\! 200 GeV.
High multiplicity $p+p$ events were those from the top 10$\%$ of
the minimum bias $p+p$ multiplicity distribution for $|\eta| \!<\!
0.5$. Since the pion daughters from $\rho^0 $ decays originate at
the interaction point, only tracks whose distance of closest
approach to the primary interaction vertex was less than 3 cm were
selected. Charged pions were selected by requiring their $dE/dx$
to be within 3 standard deviations (3$\sigma$) of the expected
mean. In order to enhance track quality \cite{16}, candidate decay
daughters were also required to have $|\eta| \!<\! 0.8$ and
transverse momenta $p_T \!>\! 0.2$ GeV/$c$.

\begin{figure}[htb]
\includegraphics[height=20pc,width=18pc]{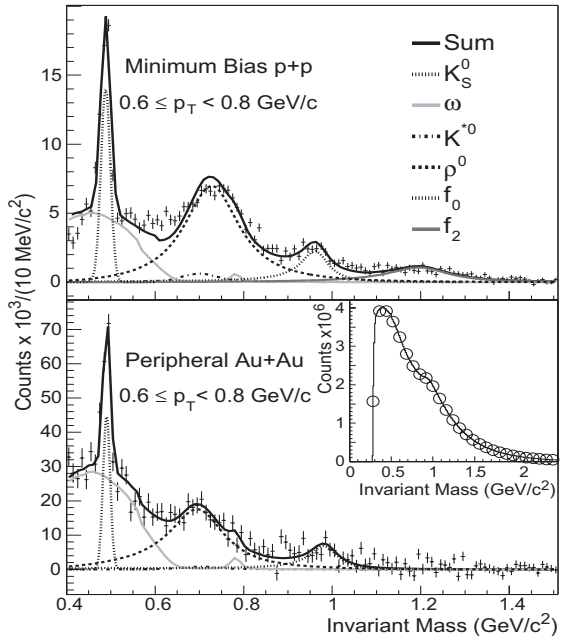}
\caption{\label{fig:cocktail}The raw $\pi^+\pi^-$ invariant mass
distributions after subtraction of the like-sign reference
distribution for minimum bias $p+p$ (top) and peripheral Au+Au
(bottom) interactions. The inset plot corresponds to the raw
$\pi^+\pi^-$ invariant mass (solid line) and the like-sign
reference distributions (open circles) for peripheral Au+Au
collisions.}
\end{figure}

The main focus of this study was the decay channel $\rho^0
\!\rightarrow\! \pi^{+}\pi^{-}$, which has a branching ratio of
$\sim$100$\%$. Similar to previous $e^+e^-$ and $p+p$
measurements, the $\rho^0$ sample studied did not select
exclusively on the $\ell$=1 $\pi^+\pi^-$ channel
\cite{17,18,19,20,21,22,23,24,25}. The measurement was performed
calculating the invariant mass for each $\pi^+ \pi^-$ pair in an
event. The resulting invariant mass distribution was then compared
to a reference distribution calculated from the geometric mean of
the invariant mass distributions obtained from uncorrelated $\pi^+
\pi^+$ and $\pi^- \pi^-$ pairs from the same events. The
$\pi^+\pi^-$ invariant mass distribution ($M_{\pi\pi}$) and the
like-sign reference distribution were normalized to each other at
$M_{\pi\pi} \gtrsim$ 1.5 GeV/$c^2$. The resulting raw
distributions for minimum bias $p+p$ and peripheral Au+Au
collisions at midrapidity ($|y| \!<\! 0.5$) for a particular $p_T$
bin are shown in Fig.~\ref{fig:cocktail}. The signal to background
is 1/10 in minimum bias $p+p$ and 1/200 in peripheral Au+Au
collisions. The $p_T$ coverage of the $\pi^+\pi^-$ pair is 0.2
$\!\leq\! p_T \!\leq\!$ 2.8 GeV/$c$ for minimum bias $p+p$ and 0.2
$\!\leq\! p_T \!\leq\!$ 2.2 GeV/$c$ for peripheral Au+Au
collisions.

The solid black line in Fig.~\ref{fig:cocktail} is the sum of all
the contributions in the hadronic cocktail. The $K_S^0$ was fit to
a Gaussian (dotted line). The $\omega$ (light grey line) and
$K^{\ast}(892)^{0}$ (dash-dotted line) shapes were obtained from
the HIJING event generator \cite{26}, with the kaon being
misidentified as a pion in the case of the $K^{\ast 0}$. The
$\rho^0(770)$ (dashed line), the $f_0(980)$ (dotted line) and the
$f_2(1270)$ (dark grey line) were fit by relativistic Breit-Wigner
functions \cite{27} BW \!=\! $M_{\pi\pi}M_0\Gamma/[(M_0^2 \!-\!
M_{\pi\pi}^2)^2 \!+\! M_0^2\Gamma^2]$ times the Boltzmann factor
\cite{5,6,7,8} PS \!=\! $(M_{\pi\pi}/\sqrt{M_{\pi\pi}^2 \!+\!
p_T^2}) \!\times\! \exp(\!-\!\sqrt{M_{\pi\pi}^2 \!+\! p_T^2}/T)$
to account for phase space. Here, $T$ is the temperature at which
the resonance is emitted \cite{6} and $\Gamma \!=\! \Gamma_0
\!\times\! (M_0/M_{\pi\pi})\!\times\![(M_{\pi\pi}^2 \!-\!
4m_\pi^2)/(M_0^2 \!-\! 4m_\pi^2)]^{(2\ell\!+\!1)/2}$ is the
momentum dependent width \cite{27}. The masses of $K_S^0$,
$\rho^0$, $f_0$, and $f_2$ were free parameters in the fit, and
the widths of $\rho^0$, $f_0$ and $f_2$ were fixed according to
\cite{28}. The uncorrected yields of $K_S^0$, $\rho^0$, $\omega$,
$f_0$, and $f_2$ were free parameters in the fit while the
$K^{\ast 0}$ fraction was fixed according to the
$K^{\ast}(892)^{0} \!\rightarrow\! \pi K$ measurement. The
$\rho^0$, $\omega$, $K^{\ast 0}$, $f_0$, and $f_2$ distributions
were corrected for the detector acceptance and efficiency
determined from a detailed simulation of the TPC response using
GEANT \cite{16}. For the particular $p_T$ bin depicted in
Fig.~\ref{fig:cocktail} and the invariant mass region shown, this
correction is approximately constant and is $\sim$25$\%$ for
minimum bias $p+p$ and varies from $\sim$25$\%$ to $\sim$35$\%$
for peripheral Au+Au collisions. The number of degrees of freedom
(d.o.f.) from the fits was 196 and the typical $\chi^2/$d.o.f. was
1.4. In the minimum bias $p+p$ invariant mass distribution shown
in Fig.~\ref{fig:cocktail}, $\pi^\pm \pi^\pm$ Bose-Einstein
correlations have been taken into account. These affect the
distribution for $M_{\pi\pi} \!<\!$ 0.45 GeV/$c^2$.

\begin{figure}[htb]
\includegraphics[height=13pc,width=19pc]{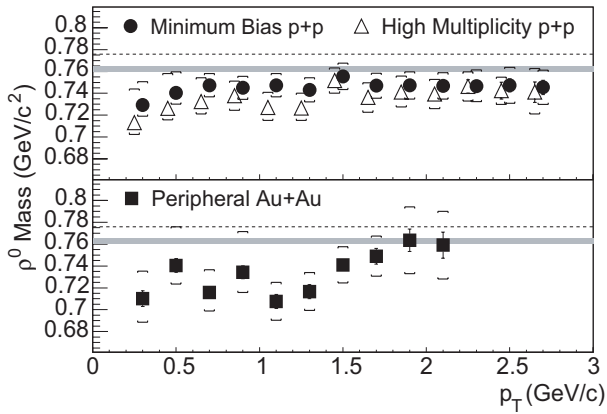}
\caption{\label{fig:mass} The $\rho^0$ mass as a function of $p_T$
for minimum bias $p+p$ (filled circles), high multiplicity $p+p$
(open triangles), and peripheral Au+Au (filled squares)
collisions. The error bars indicate the systematic uncertainty.
Statistical errors are negligible. The $\rho^0$ mass was obtained
by fitting the data to the BW$\times$PS functional form described
in the text. The dashed lines represent the average of the
$\rho^0$ mass measured in $e^+e^-$ \cite{28}. The shaded areas
indicate the $\rho^0$ mass measured in $p+p$ collisions \cite{22}.
The open triangles have been shifted downward on the abscissa by
50 MeV/$c$ for clarity.}
\end{figure}

The $\rho^0$ mass is shown as a function of $p_T$ in
Fig.~\ref{fig:mass} for peripheral Au+Au, high multiplicity $p+p$,
and minimum bias $p+p$ interactions. The $\rho^0$ mass was
obtained by fitting the data to a relativistic $p$-wave ($\ell$
\!=\! 1) Breit-Wigner function times a factor which accounts for
phase space (BW$\times$PS) in the hadronic cocktail. Since the
phase space factor modifies the position of the peak for the BW
function, the mass derived from the BW$\times$PS fit may be
shifted compared to the peak of the experimental invariant mass
distribution and to the peak of the BW function alone. The
$\rho^0$ peak was also fit to a relativistic $p$-wave BW function
excluding the PS factor in the hadronic cocktail; however, the fit
failed to reproduce the $\rho^0$ line shape, and underestimated
the position of the peak in general, particularly at low $p_T$.
This measurement does not have sufficient sensitivity to permit a
systematic study of the $\rho^0$ width. Therefore, for the
cocktail fits in this analysis, the $\rho^0$ width was fixed at
$\Gamma_0$ \!=\! 160 MeV$/c^2$, consistent with folding the
$\rho^0$ natural width (150.9 $\!\pm\!$ 2.0 MeV$/c^2$ \cite{28})
with the intrinsic resolution of the detector \cite{16}. In Au+Au
collisions, the temperature used in the PS factor was $T$ \!=\!
120 MeV \cite{6}, while in $p+p$, $T$ \!=\! 160 MeV \cite{29}.

The $\rho^0$ mass at $|y| \!<\!$ 0.5 for minimum bias $p+p$, high
multiplicity $p+p$, and peripheral Au+Au collisions at $\sqrt{s}$
$\!=\!$ 200 GeV seems to increase as a function of $p_T$ and is
systematically lower than the value reported by \cite{22}. The
$\rho^0$ mass measured in peripheral Au+Au collisions is lower
than the minimum bias $p+p$ measurement. The $\rho^0$ mass for
high multiplicity $p+p$ interactions is lower than for minimum
bias $p+p$ interactions for all $p_T$ bins, showing that the
$\rho^0$ mass is also multiplicity dependent. Recent calculations
are not able to reproduce the $\rho^0$ mass measured in peripheral
Au+Au collisions without introducing in-medium modification of the
$\rho^0$ meson \cite{6,7,8,9,10,33}.

Previous observations of the $\rho$ meson in $e^+e^-$
\cite{30,31,32} and $p+p$ interactions \cite{22} indicate that the
$\rho^0$ line shape is considerably distorted from a $p$-wave BW
function. A mass shift of $-$30 MeV/$c^{2}$ or larger was observed
in $e^+e^-$ collisions at $\sqrt{s}$ \!=\! 90 GeV \cite{30,31,32}.
In the $p+p$ measurement at $\sqrt{s}$ \!=\! 27.5 GeV \cite{22}, a
$\rho^0$ mass of 0.7626 $\!\pm\!$ 0.0026 GeV/$c^2$ was obtained
from a fit to the BW$\times$PS function \cite{11,22}. However, in
this measurement the position of the $\rho^0$ peak is lower than
the average of the $\rho^0$ mass measured in $e^+e^-$ interactions
\cite{28} by $\sim$30 MeV/$c^2$ \cite{22}. This result is the only
$p+p$ measurement used in the hadroproduced $\rho^0$ mass average
reported in \cite{28}.

In comparison to the in-medium $\rho^0$ production in hadronic
Au+Au interactions, no modifications of the $\rho^0$ properties
are expected for coherent $\rho^0$ production in ultraperipheral
heavy-ion collisions, where (in lowest order) at impact parameters
$b \!>\! 2 R_A$, a photon emitted by one gold ion fluctuates into
a virtual $\rho^0$ meson state, which scatters diffractively from
the other nucleus. The $\rho^0$ line shape in ultra-peripheral
collisions measured  with the STAR detector \cite{27} is
reproduced by a BW plus S\"oding interference term, with the
$\rho^0$ mass and width consistent with their natural values
reported in \cite{28}.

One uncertainty in the hadronic cocktail fit depicted in
Fig.~\ref{fig:cocktail} is the possible existence of correlations
of unknown origin near the $\rho^0$ mass. An example is
correlations in the invariant mass distribution from particles
like the $f_0(600)$ which are not well established \cite{28}. The
$\omega$ yield in the hadronic cocktail fits may account for some
of these contributions and may cause the apparent decrease in the
$\rho^0/\omega$ ratio between minimum bias $p+p$ and peripheral
Au+Au interactions. In order to evaluate the systematic
uncertainty in the $\rho^0$ mass due to poorly known contributions
in the hadronic cocktail, the $\rho^0$ mass was obtained by
fitting the peak to the BW$\times$PS function plus an exponential
function representing these contributions. Using this procedure,
the $\rho^0$ mass is systematically higher than the mass obtained
from the hadronic cocktail fit. This uncertainty is the main
contribution to the systematic uncertainties shown in
Fig.~\ref{fig:mass} and it can be as large as $\sim$35 MeV/$c^2$
for low $p_T$. Other contributions to the systematic errors shown
in Fig.~\ref{fig:mass} result from uncertainty in the measurement
of particle momenta of $\sim$3 MeV/$c^2$ (this leads to a mass
resolution of $\sim$8 MeV/$c^2$ at the $\rho^0$ mass) and from the
hadronic cocktail fits themselves of $\sim$13 MeV/$c^2$. The
systematic uncertainties are common to all $p_T$ bins and are
correlated between the $p+p$ and peripheral Au+Au measurements.

\begin{figure}[htb]
\includegraphics[height=12pc,width=18pc]{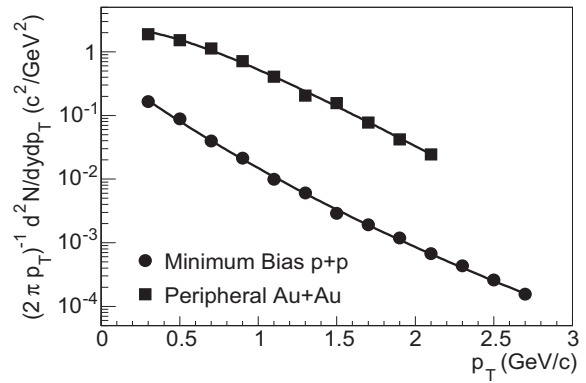}
\caption{\label{fig:spectra} The $p_T$ distributions at $|y|
\!<\!$ 0.5 for minimum bias $p+p$ and peripheral Au+Au collisions.
See text for an explanation of the functions used to fit the data.
The errors shown are statistical only and smaller than the
symbols.}
\end{figure}

The corrected invariant yields [$d^2N/(2\pi p_Tdp_Tdy)$] at $|y|
\!<\!$ 0.5 as a function of $p_T$ for peripheral Au+Au and minimum
bias $p+p$ interactions are shown in Fig.~\ref{fig:spectra}. In
$p+p$ interactions, a power-law fit was used to extract the
$\rho^0$ yield per unit of rapidity around midrapidity. The fit
yielded $dN/dy$ \!=\! 0.259 $\!\pm\!$ 0.002(stat) $\!\pm\!$
0.039(syst) and $\langle p_T \rangle \!=\!$ 0.616 $\!\pm\!$
0.002(stat) $\!\pm\!$ 0.062(syst) GeV/$c$. In Au+Au collisions, an
exponential fit in $m_T - m_0$, where $m_0 \!=\! 0.769$ MeV/$c^2$
is the average $\rho^0$ mass reported in \cite{28}, was used to
extract the $\rho^0$ yield and the inverse slope. The fit yielded
$dN/dy$ \!=\! 5.4 $\!\pm\!$ 0.1(stat) $\!\pm\!$ 1.2(syst) and an
inverse slope of 318 $\!\pm\!$ 4(stat) $\!\pm\!$ 38(syst) MeV
[$\langle p_T \rangle \!=\!$ 0.83 $\!\pm\!$ 0.01(stat) $\!\pm\!$
0.10(syst) GeV/$c$]. The main contributions to the systematic
uncertainties quoted are due to the tracking efficiency
($\sim$8$\%$) and the normalization between the $M_{\pi\pi}$ and
the like-sign reference distributions ($\sim$9$\%$ for minimum
bias $p+p$ and $\sim$19$\%$ for peripheral Au+Au collisions).

\begin{figure}[htb]
\includegraphics[height=13pc,width=18pc]{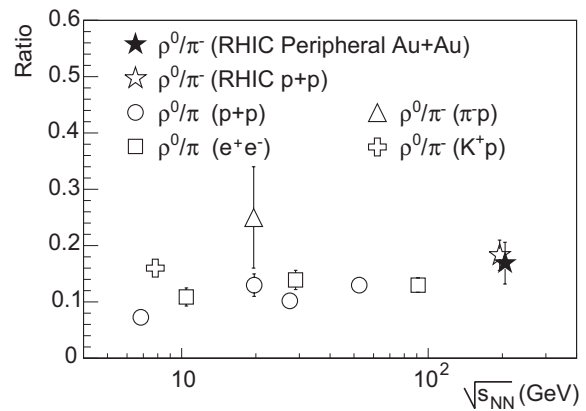}
\caption{\label{fig:ratios} $\rho^0/\pi$ ratios as a function of
c.m. system energy. The ratios are from measurements in $e^+e^-$
collisions at 10.45 GeV \cite{17}, 29 GeV \cite{18} and 91 GeV
\cite{19} c.m. system energy, $p+p$ at 6.8 GeV \cite{20}, 19.7 GeV
\cite{21}, 27.5 GeV \cite{22}, and 52.5 GeV \cite{23}, $K^+p$ at
7.82 GeV \cite{24} and $\pi^-p$ at 19.6 GeV \cite{25}. The errors
on the ratios at $\sqrt{s_{_{NN}}}$\!=\! 200 GeV are the quadratic
sum of the statistical and systematic errors. The ratios at
$\sqrt{s_{_{NN}}}$\!=\! 200 GeV are offset from one another for
clarity.}
\end{figure}

The $\rho^0/\pi^-$ ratio is 0.183 $\!\pm\!$ 0.001(stat) $\!\pm\!$
0.027(syst) for minimum bias $p+p$, and 0.169 $\!\pm\!$
0.003(stat) $\!\pm\!$ 0.037(syst) for peripheral Au+Au collisions.
The comparison with measurements in $e^+e^-$ \cite{17,18,19},
$p+p$ \cite{20,21,22,23}, $K^+p$ \cite{24}, and $\pi^-p$ \cite{25}
interactions at different c.m. system energies is shown in
Fig.~\ref{fig:ratios}. The $\rho^0/\pi^-$ ratios from minimum bias
$p+p$ and peripheral Au+Au interactions are comparable.

The $\rho^0/\pi^-$ ratios from statistical model calculations
\cite{8,9,13} for Au+Au collisions are considerably lower than the
measurement presented in Fig.~\ref{fig:ratios}. The larger
$\rho^0/\pi^-$ ratio measured may be due to the interplay between
the rescattering of the $\rho^0$ decay products and $\rho^0$
regeneration.

In conclusion, we have presented results on $\rho(770)^0$
production at midrapidity in minimum bias $p+p$ and peripheral
Au+Au collisions at $\sqrt{s_{_{NN}}}$ \!=\! 200 GeV. This is the
first direct measurement of $\rho^0(770) \!\rightarrow\!
\pi^+\pi^-$ in heavy-ion collisions. The $\rho^0$ mass seems to
increase slightly as a function of $p_T$, and to decrease with
multiplicity. The measured $\rho^0$ peak in the invariant mass
distribution is lower than previous measurements reported in
\cite{28} by $\sim$40 MeV/$c^2$ in minimum bias $p+p$ interactions
and $\sim$70 MeV/$c^2$ in peripheral Au+Au collisions. Similar
mass shifts were observed in $e^+e^-$ and $p+p$ interactions.
Dynamical interactions with the surrounding matter, interference
between various $\pi^+\pi^-$ scattering channels, phase space
distortions due to the rescattering of pions forming $\rho^0$, and
Bose-Einstein correlations between $\rho^0$ decay daughters and
pions in the surrounding matter are possible explanations for the
apparent modification of the $\rho^0$ meson properties. The
$\rho^0/\pi^-$ ratio in peripheral Au+Au collisions is higher than
predicted by statistical calculations, and is comparable to the
measured value in minimum bias $p+p$ interactions. Further
measurements of the $\rho^0$ meson, along with other resonance
particles, can provide important information on the dynamics of
relativistic collisions and help in understanding the properties
of nuclear matter under extreme conditions.

We thank M. Bleicher, P. Braun-Munzinger, W. Broniowski, G.E.
Brown, W. Florkowski, P. Kolb, G.D. Lafferty, S. Pratt, R. Rapp,
and E. Shuryak for valuable discussions. We thank the RHIC
Operations Group and RCF at BNL, and the NERSC Center at LBNL for
their support. This work was supported in part by the HENP
Divisions of the Office of Science of the U.S. DOE; the U.S. NSF;
the BMBF of Germany; IN2P3, RA, RPL, and EMN of France; EPSRC of
the United Kingdom; FAPESP of Brazil; the Russian Ministry of
Science and Technology; the Ministry of Education and the NNSFC of
China; SFOM of the Czech Republic, FOM and UU of the Netherlands,
DAE, DST, and CSIR of the Government of India; the Swiss NSF.

\end{document}